\documentclass[twocolumn]{aastex631}

\usepackage{CJKutf8}
\usepackage{color}

\usepackage{graphicx}	
\usepackage{amsmath}	
\usepackage{amssymb}	

\newcommand\lsim{\mathrel{\rlap{\lower4pt\hbox{\hskip1pt$\sim$}}
\raise1pt\hbox{$<$}}}
\newcommand\gsim{\mathrel{\rlap{\lower4pt\hbox{\hskip1pt$\sim$}}
\raise1pt\hbox{$>$}}}

\shortauthors{Faridani et al.}

\graphicspath{{./}{figures/}}

\begin{document}
\title{Hiding Planets Near and Far: The Parameter Space of Hidden Companions for Known Planetary Systems}
\shorttitle{Hiding Planets Near and Far}

\begin{CJK*}{UTF8}{gbsn}

\author[0000-0003-3799-3635]{Thea H. Faridani}
\correspondingauthor{Thea H. Faridani}
\email{thfaridani@astro.ucla.edu}
\affil{Department of Physics and Astronomy, University of California, Los Angeles, CA 90095, USA\\}
\affil{Mani L. Bhaumik Institute for Theoretical Physics, Department of Physics and Astronomy, UCLA, Los Angeles, CA 90095, USA\\}

\author[0000-0002-9802-9279]{Smadar Naoz}
\affil{Department of Physics and Astronomy, University of California, Los Angeles, CA 90095, USA\\}
\affil{Mani L. Bhaumik Institute for Theoretical Physics, Department of Physics and Astronomy, UCLA, Los Angeles, CA 90095, USA\\}

\author[0000-0002-2612-2933]{Lingfeng Wei (魏凌枫)}
\affiliation{Center for Astrophysics and Space Sciences, University of California, San Diego, La Jolla, CA 92093, USA}

\author[0000-0003-1540-8562]{Will M. Farr}
\affiliation{Department of Physics and Astronomy, Stony Brook University, Stony Brook NY 11794, USA}
\affiliation{Center for Computational Astrophysics, Flatiron Institute, 162 Fifth Avenue, New York, NY 10010, USA}

\begin{abstract}

Recent ground and space-based observations show that stars with multiple planets are common in the galaxy. Most of these observational methods are biased toward detecting large planets near to their host stars. Because of these observational biases, these systems can hide small, close-in planets or far-orbiting (big or small) companions. These planets can still exert dynamical influence on known planets and have such influence exerted upon them in turn. In certain configurations, this influence can destabilize the system; in others, the star's gravitational influence can instead further stabilize the system. For example, in systems with planets close to the host star, effects arising from general relativity can help to stabilize the configuration.  We derive criteria for hidden planets orbiting both beyond and within known planets that quantify how strongly general relativistic effects can stabilize systems that would otherwise be unstable. As a proof-of-concept, we investigate the several planets {in a system based on Kepler 56, and} show that the outermost planet will not disrupt the system {even at high eccentricities}, and show that an Earth-radius planet could be stable within this system if it orbits below $0.08$ au. Furthermore, we provide specific predictions to known observed systems by constraining the parameter space of possible hidden planets.

\end{abstract}


\section{Introduction}

Recent developments in ground and space based observations have found that multi-planet systems are common in the Galaxy \citep[e.g.,][]{Howard+10,Howard+12,Lissauer+11}.
Current leading exoplanet detection methods are are more sensitive to larger planets and to planets with shorter periods. For example, the popular transit
method is particularly sensitive to close-in planets \citep[]{Borucki+16}. 

Because the transit method represents on the order of three quarters of all current exoplanet detections, the biases inherent in the method leave open two classes of planets that are currently difficult to detect: small planets close to their hosts and large planets further out. Radial velocity detection methods, as the second most prolific method, give access to large regimes of wide-orbit parameter space, and have shed light on the life cycle of giant planets living there \citep[]{Hatzes+05, Knutson+14K, Stock+18}. 

The observed statistical properties of  multi-planet systems  have been used to infer the underlying history and evolution of planets \citep[e.g.,][]{Fang+12,Hansen+13,Malhotra+15,Pu+15,Steffen+15,Ballard+16,Xie+16,Weiss+18Peas,Pu+18,Denham+19, Tamayo+20}. 
One curious property of the multiplanet systems that have been observed is that the radii of planets in the same system are correlated with one another \citep[][]{Weiss+18Peas,Weiss+18multis}. There {was} debate whether this has been conclusively demonstrated to arise from formation processes as opposed to observation biases \citep[e.g.,][]{Zhu20,Weiss_2020,2020AJ....160..160M}, considering that dynamical evolution may erase the initial conditions of formation, {however, correlations not explainable by observation bias have been demonstrated} \citep[e.g.,][]{He+21,gilbert+20}. 
Therefore, when we consider hypothetical configurations of planets that are meant to be similar to configurations already observed, we usually assume correlated planet masses.

Observations suggest that far-away companions (giant planets or even stars) are expected to host inner multi-planet systems.  This trend was suggested by radial velocity surveys that showed a population of giant planets may exist at large distances from stars that host one or more inner planets \citep[e.g.,][]{Knutson+14K,Konopacky+2016,Zhu+18,Bryan+16,Bryan+19}. Additionally, several systems have been discovered where the configuration is strictly hierarchical, with a giant planet on a wide orbit outside of one or more very close-in planets \citep[e.g.,][]{2017A&A...602A.107B,2018AJ....156..213M,2018A&A...615A.175B,Mills+19, Bryan+19, Zhu+18}. Furthermore, the most generic example is our solar system, with Jupiter at $5$~au. Lastly, since the majority of stars exist in a binary configuration \citep[e.g.,][]{Raghavan+10,MoeStefano+15}, {a population of distant stellar companions to multi-planet systems is to be expected.}

Companions that may exist far away from their host stars may dynamically affect an inner planetary system.
For example, gravitational perturbations from a far-away companion (either a stellar or planetary companion) can excite the inner planets' eccentricity, via the Eccentric Kozai-Lidov (EKL) mechanism \citep[e.g.,][]{Kozai,Lidov,Naoz16}. High eccentricity, in combination with tidal dissipation, can lead to shrinking of the semi-major axis of a planet, circularizing it \citep[e.g.,][]{Fabrycky+07,Wu+07,Naoz11,Naoz+12bin,Li+14coplaner,Rice15,Petrovich15HJ,Lai+18,Stephan+18,martin+15}. On the other hand, the eccentricity excitations may plunge the planet into the star \citep[e.g.,][]{Naoz+12bin,Petrovich15,Lai+18,Stephan+18,Stephan+20}. In some cases, short-range forces, such as general relativity and tidal precession may suppress these eccentricity excitations \citep[e.g.,][]{Fabrycky+07,Naoz+13,Liu+15,Mardling+04,Plavchan+15}. In general,  during the formation of planets, Jupiter-size planets have a large role in the final structure of the system, by driving planetesimals inward  \citep[e.g.,][]{Morbidelli+07, raymond+17} or by controlling the ability of planetesimals to accrete \citep[e.g.,][]{Hansen+09,walsh+11}.

In the case of a multi-planet system, high eccentricity may result in destabilization. However, the gravitational interactions between close-by multi-planetary orbits, and in particular, the angular momentum exchange between different planets, may suppress high eccentricity induced by a far-away companion \citep[e.g.,][]{Inn+97,Li+14Kepler,Hansen17,Pu+18,Denham+19,wei+21,Veras+10, Holman+97,Naoz+12GR}. Additionally, it may result in synchronized oscillation of the inclinations of all inner planets, resulting in a large obliquity---an observational marker that there may be a hidden far-away companion \citep[e.g.,][]{Inn+97,Takeda05,Takeda,Li+14Kepler,Denham+19,boue+14}.

On the other extreme, the lower mass limit of transit surveys today lies on the order of half an Earth radius, for a $1$~M$_\odot$ host star \citep[assuming a {favorable} noise level][]{2012PASP..124.1279C}.
These short-period planets, usually said to have "ultra short periods" (USPs) of less than one day, produce so many transits that they can be detected despite their small size. {Already, over a hundred such planets have been  detected} \citep[e.g.,][]{2014ApJ...787...47S, 2020ChA&A..44..283X,2017AJ....154..226D}. One of the key observed qualities of USPs is that they may have radii $\lsim R_\oplus$. Their small radii have been hypothesized to have been linked to the large irradiances they receive \citep[e.g.,][]{2020A&A...633A.133F}. Currently, the properties of the atmospheres of USPs (if they have any atmosphere at all) are relatively unknown \citep{2018AJ....155..107M}. Because some models explaining the super earth radius gap rely on strong stellar flux incoming to the star \citep[e.g.,][]{Owen+16,Owen+17,2020AJ....160...89P,2020ApJ...890...23L}\footnote{We note that there are alternatives to the photo-evaporation model \citep[see][]{Gupta+20,Gupta+21}.}, USPs {may be} ideal candidates for constraining these models. 
Similarly, a range of formation scenarios exist that can explain how these planets came to have such short periods, such as secular chaos and disk migration \citep[e.g.,][]{2019MNRAS.488.3568P,2019MNRAS.486.3874C,2019AJ....157..180P,Becker+20}. Though most models predict that they do not form completely \textit{in situ}, at what stage in their evolution they are brought close to the star remains unclear \citep[e.g.,][]{2020ApJ...905...71M}.%

In this work, we examine the stability of planets in the radius and orbit ranges where their detection via the transit method is currently not possible. Historically, the techniques used in this work have exclusively predicted the permitted locations of large, distant companion planets \citep[e.g.,][]{Denham+19,Pu+18}. We extend these {approaches} by incorporating non-Newtonian forces arising from General Relativity (GR), allowing for constraints to be placed on short-period planets as well. {In particular, we demonstrate, as a proof of concept, the usage of this analytical stability criterion for several hypothetical and observational examples.}

This paper is structured as follows. In Section \ref{sec:eqs} we introduce the basic equations used throughout the analysis and show their use in quantifying the stability of systems. We then focus on systems with known two or more planets and put constraints on which of those systems have their stability against perturbation from an outer companion extended by GR precession  (see Section \ref{sec:OuterGR}). We then turn our attention to possible ultra-short period planets (USPs) (Section \ref{sec:Inner}), and explore the role of GR for these systems (see for the derivation of an analytical stability criterion in Section \ref{sec:GRInner}). GR precession of USPs tends to suppress eccentricity excitations that may be induced due to Laplace-Lagrange resonances (as shown in Section \ref{sec:LLGR}. We then show that our USPs' stability analytical criterion agrees compared to numerical direct $n$-body integration (see Section \ref{sec:num}).

In Section \ref{sec:Kepler56}, as a case study, we examine the system Kepler 56 and explore the interaction between general relativity and traditional secular methods within it. Finally, in Section \ref{sec:diss} we discuss our findings.

\section{The Two Known Planets - Basic Equations}\label{sec:eqs}

\begin{figure}
    \includegraphics[width=\linewidth]{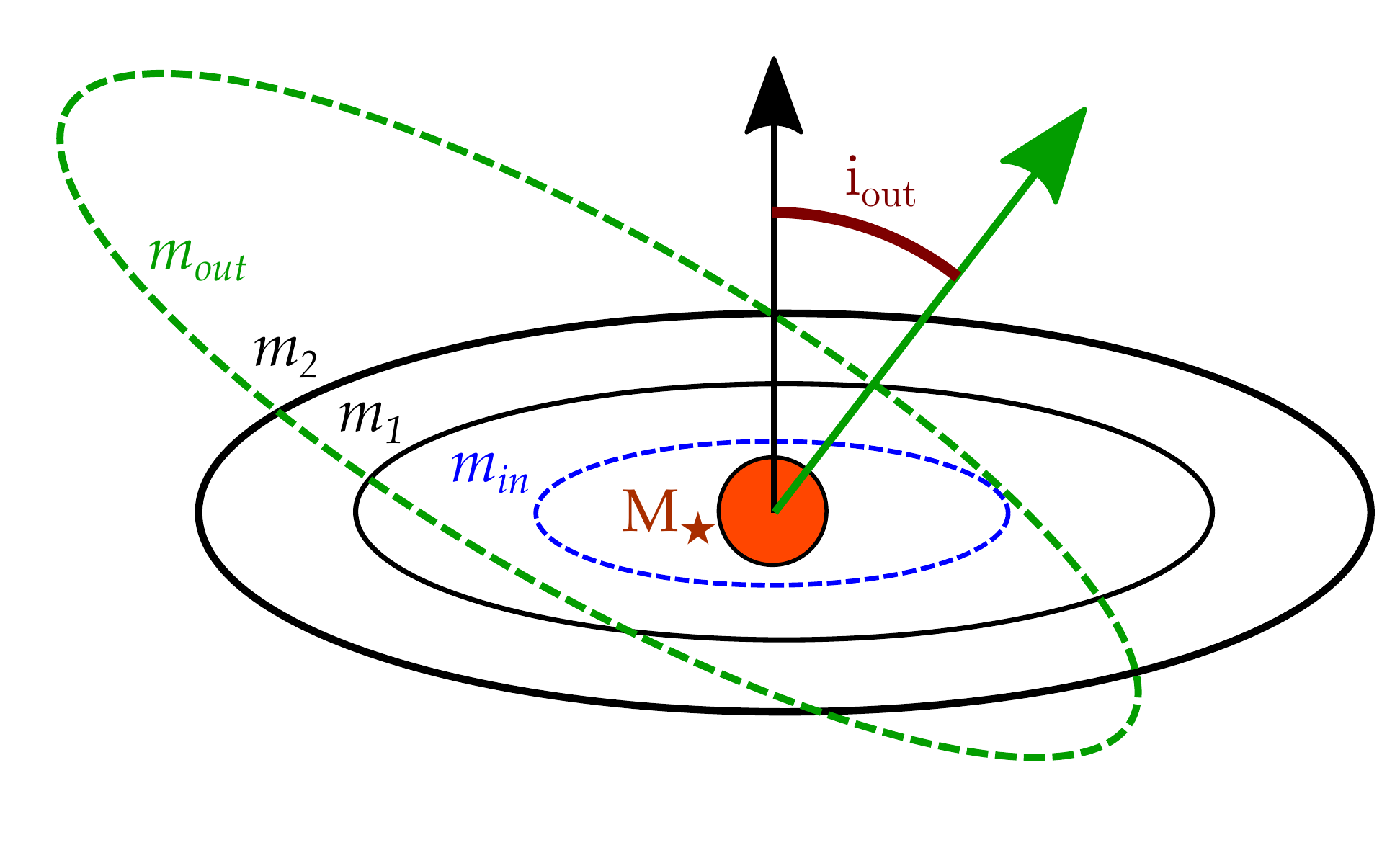}
    \caption{Schematic diagram (not to scale) of the systems discussed in this paper. A star of mass $M_\star$ has one or two known planets, $m_1$ and $m_2$. We will consider the stability of this system after the insertion of a hypothetical inner planet $m_{\rm in}$, or a hypothetical outer planet $m_{\rm out}$.}
    \label{fig:cartoon}
\end{figure}
    
We consider a fiducial system composed of a star $M_\star$ with two observed planets, $m_1$ and $m_2$, and explore the possibility that a planet may be hiding undetected inward ($m_{\rm in}$) or outward ($m_{\rm out}$) of the known planets (see Figure \ref{fig:cartoon} for an illustration of the system). The known planets have semi-major axes (eccentricities), denoted by  $a_1$ and $a_2$ ($e_1$ and $e_2$) where $a_1<a_2$.  Angular momentum exchange between the {inner} planets may stabilize them against gravitational perturbations from the outer companion \citep[e.g.,][]{Denham+19}. Additionally, general relativity precession of the periapsis may instead further stabilize the system{, including a possible inner planet from the perturbations of its neighbors.} 

Here we describe the basic equations that govern the system. The exchange of angular momentum between planets planets in the systems is described via the Laplace-Lagrange formalism. Given planets with  masses $m_j$ and $m_k$, the characteristic timescale over which the angular momentum  of planet $j$ changes due to planet $k$, can be written as: 

\begin{align}\label{eq:tau_LL_full}
    \tau_{jk, {\rm LL}} = \Bigg[ & A_{jj} +  A_{jk}\left(\frac{e_k}{e_j}\right)\cos\left(\varpi_k-\varpi_j\right) \nonumber \\
    - & B_{jj} - B_{jk}\left(\frac{i_k}{i_j}\right)\cos\left(\Omega_k - \Omega_j\right)\Bigg]^{-1} \ ,
\end{align}
where $e_j$ ($e_k$) is the eccentricity, and $i_j$ ($i_j$) is the inclination of planet $j$ ($k$) and $\Omega_j$ ($\Omega_k$) is the longitude  of ascending nodes of planet $j$ ($k$), while  $\varpi_j$ ($\varpi_k$) is the longitude of periapsis. The coefficients $A_{jk}$ and $B_{jk}$ are given by \citep[e.g.,][]{Murray+00book}:
\begin{equation}
    A_{jj} = \frac{n_j m_k}{4\pi \left( M+m_j\right)} \alpha_{jk} \overline{\alpha}_{jk} f_\psi  \ ,
    \label{eq:A_matrix}
\end{equation}
\begin{equation}
    A_{jk} = -\frac{n_j m_k}{4\pi \left( M+m_j\right)} \alpha_{jk} \overline{\alpha}_{jk} f_{2\psi}  \ ,
\end{equation}
\begin{equation}
    B_{jj} = -\frac{n_j m_k}{4\pi \left( M+m_j\right)} \alpha_{jk} \overline{\alpha}_{jk} f_\psi  \ ,
\end{equation}
\begin{equation}
    B_{jk} = \frac{n_j m_k}{4\pi \left( M+m_j\right)} \alpha_{jk} \overline{\alpha}_{jk} f_\psi  \ ,
\end{equation}
where,
\begin{equation}
    \alpha_{jk} = \min \left(\frac{a_j}{a_k},\frac{a_k}{a_j}  \right) \ ,
\end{equation}
\begin{equation}
    \overline{\alpha}_{jk} = \min \left(\frac{a_j}{a_k},1  \right) \ ,
\end{equation}
and 
\begin{equation}
f_{\psi}=\int_{0}^{2 \pi} \frac{\cos \psi}{\left(\alpha_{j k}^{2}-2 \alpha_{j k} \cos \psi+1\right)^{3 / 2}} d \psi \ ,
\end{equation}
\begin{equation}
f_{2\psi}=\int_{0}^{2 \pi} \frac{\cos 2\psi}{\left(\alpha_{j k}^{2}-2 \alpha_{j k} \cos \psi+1\right)^{3 / 2}} d \psi \ .
\end{equation}

We assume that the two known planets were born \textit{in situ}, i.e., their eccentricities and mutual inclinations are small and similar in value. {We note that this not a universally advisable assumption.} Under this assumption, the precession of the orbit of planet $j$ due to the aforementioned angular momentum exchange with planet $k$ takes place over a characteristic timescale, estimated as: 
\begin{equation}
    \tau_{LL,jk} = \left[ \frac{n_j m_k}{4\pi (M_\star + m_k)}\alpha_{jk}\overline{\alpha}_{jk} \left(f_{\psi}\pm{f_{2\psi}} \right) \right]^{-1} \ ,
    \label{eq:tau_LL_pm}
\end{equation}
where $n_j$ is the mean motion of planet $j$, and the two cases $\left(f_{\psi}\pm{f_{2\psi}} \right)$ correspond to the the minimum ($+$) and maximum ($-$) values this timescale can achieve. 

Neither $f_{\psi}$ nor $f_{2\psi}$ have a known closed-form representation to our knowledge, but under the assumption that $\alpha_{jk} \leq 0.9$, we find
\begin{equation}\label{eq:approx}
    \left(f_{\psi}-{f_{2\psi}} \right) \approx 7\alpha_{jk} \ ,
\end{equation}
to within an error of not more than $40\%$. Because this term has a simple closed-form approximation, when deriving our criterion, we use the $\left(f_{\psi}-{f_{2\psi}} \right)$ case over the $\left(f_{\psi}+{f_{2\psi}} \right)$

In addition to the Newtonian effect between the two planets, general relativity produces a precession of their orbits. {Thus, a far away companion may drive high eccentricity on the inner planets, but in addition to stabilization from Laplace-Lagrange interaction, general relativity precession may suppress this eccentricity excitation} \citep[][]{wei+21}. The timescale of apsidal precession on planet $j$ caused by general relativity is given by \citep[e.g.,][]{Naoz+12GR}
\begin{equation}\label{eq:tGR}
    \tau_{GR, j} = 2 \pi \frac{c^2 a_j^{5/2} (1-e_j^2)}{3 (GM_\star)^{3/2}} \ ,
\end{equation}
where $c$ is the speed of light.
%

\section{An Outer Companion: The Role of GR for Known Systems}\label{sec:OuterGR} 
Consider an outer, inclined companion
($m_{\rm out}$) to two inner planets ($m_1$, $m_2$) that satisfy $a_1<a_2$. Including this outer companion makes the system hierarchical. This outer companion may excite the eccentricities of the inner planets through the EKL mechanism \citep[e.g.,][]{Naoz16}. The associated timescale for this eccentricity excitation on an inner planet $j$ is given by \citep[e.g.,][]{Antognini15}
\begin{equation}
\tau_{\mathrm{EKL,j}} = \frac{16}{15} \frac{a_{\rm out}^{3}\left(1-e_{\rm out}^{2}\right)^{3 / 2} \sqrt{\left(M_\star +m_{j}\right) / G}}{a_{j}^{3 / 2} m_{\rm out}} \ ,
\end{equation}
where  $a_{\rm out}$ is the outer companion semi-major axis. On the other hand, Laplace-Lagrange interactions, which exchange angular momentum between the inner planets, can suppress the outer companion's induced eccentricity excitations \citep[e.g.,][]{Inn+97,Takeda05,Takeda,Li+14Kepler,Pu+18,Denham+19}. For the suppression to occur, the angular momentum exchange due to Laplace-Lagrange interactions, needs to take place on a shorter timescale than the EKL typical timescale. The timescale of the Laplace-Lagrange effect on  planet $2$'s orbit from the gravity of planet $1$ is approximately (see Equation (\ref{eq:tau_LL_full})): 

\begin{equation}\label{eq:tauLLmax}
    \tau_{\rm LL,21, max} \approx \left[\frac{7 n_2 m_1}{4\pi M_\star} \left( \frac{a_1}{a_2} \right)^2 \right]^{-1} \ .
\end{equation}
\citet{Denham+19} {derive an analytical stability criterion indicating the region of parameter space in which a two-planet system is stable against gravitational perturbations from a far away companion.} However, short range forces can also tend to stabilize a system \citep[this was noted in the three body systems, e.g.,][]{Ford00Pls,Naoz+12GR,Liu+15,Hansen+20}. 
Recently \citet{wei+21} generalized the aforementioned stability criterion to include general short range forces. In particular, here we consider general relativity precession due to the star\footnote{Tidal precession is also a relevant short-range force that may work to stabilize the system. General relativity precession is often more important compared to tidal precession \citep[e.g.,][]{Dan,Liu+15}.}.

As was shown in \citet{wei+21} general relativity can only extend the stability regime beyond what was shown by \citet{Denham+19} if the GR timescale is shorter than the Laplace-Lagrange timescale between the two inner planets.
In other words, if the timescale hierarchy is such that $\tau_{\rm GR, 2}<\tau_{\rm EKL, 2}$, but $\tau_{\rm GR, 2}>\tau_{\rm LL, 21}$ then the system will be stable against EKL excitation by the outer perturber \citep[as shown by][]{Denham+19}, but may still be susceptible to detail-dependent excitations by Laplace-Lagrange, as we will discuss in section \ref{sec:Inner}.

Considering the population of known two-body systems, we ask in which systems GR effects further help stabilize the system more than the stabilization achieved by Laplace-Lagrange. {Observed two-body systems are likely to be stable (or marginally stable) because they have survived long enough for us to observe them.} In such a system, if there is an outer perturber, the outermost known planet is affected by the perturber's EKL excitations more than the inner known planet. Thus, if a hidden companion exists, and GR is assumed not to be relevant, the timescale hierarchy is $\tau_{\rm EKL, 2}>\tau_{\rm LL, 21}$. Therefore, to examine the importance of GR effects, we must ask when does GR stabilize $m_2$ against EKL more than Laplace-Lagrange does, i.e., when EKL excitation is involved, GR is necessary to include only when $\tau_{\rm GR,2} < \tau_{\rm LL, 21, max}$, which produces the criterion:
\begin{equation}\label{eq:criterion}
\frac{a_1^2}{a_2} \leq \frac{6 G M_\star^2}{7 c^2 (1-e_{\rm max}^2) m_1},
\end{equation}
where we used the approximation in Equation (\ref{eq:approx}). {Using $\tau_{\rm LL, 21, min}$, we do not have a neat approximation for $\left(f_{\psi}+{f_{2\psi}} \right)$, so the criterion becomes}
\begin{equation}\label{eq:criterion_min}
a_1 \left(f_{\psi}+{f_{2\psi}} \right) \leq \frac{6 G M_\star^2}{c^2 (1-e_{\rm max}^2) m_1}.
\end{equation}

\begin{figure}
	\includegraphics[width=\columnwidth]{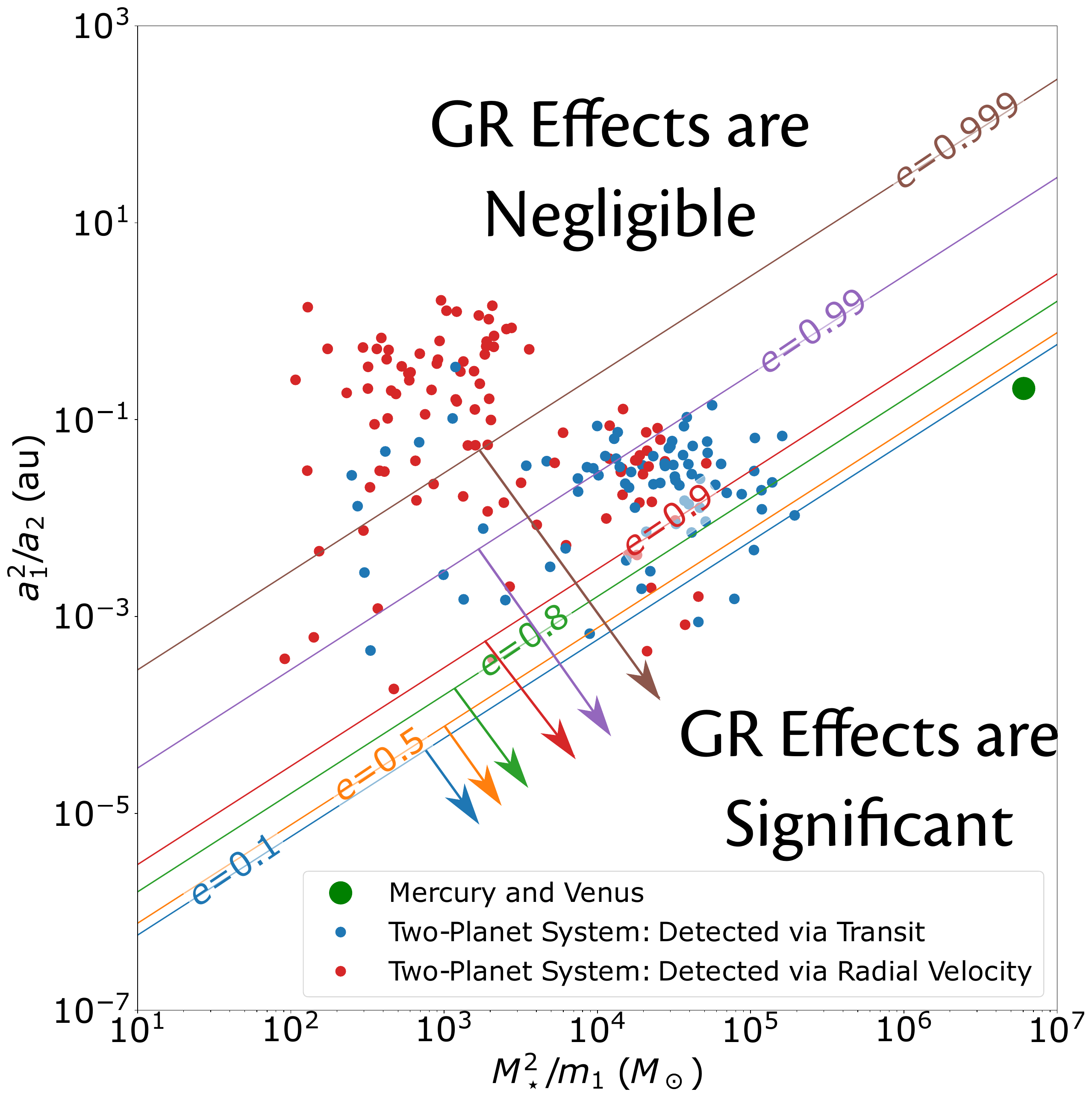}
    \caption{ Susceptibility of known, two-planet systems to GR necessity. {Systems without measured stellar masses are not included.} Known two-planet systems (blue{ and red for planets detected by the Transit Method and the Radial Velocity method, respectively}) in the parameter space defined by Equation (\ref{eq:criterion}), which defines the ${a_1^2}/{a_2}$\,-\,${M^2}/{m}$ plane.
    Overplotted are lines for different $e_{\rm max}$ that can be induced by EKL. Below the lines  $\tau_{\rm GR}(m_2) < \tau_{\rm LL}$. 
    Plotted in green is the Mercury-Venus pair. Systems lying below a given curve have their stability against EKL excitation extended by GR. Notably, At moderate-to-large values $e_{\rm max}$, the inclusion of GR extends the stability regime of a significant number of systems.} 
    
    \label{fig:grsuscept}
\end{figure}

{In these equations, $e_{\rm max}$ represents the maximum eccentricity induced by the quadrupole-level, EKL oscillation in the presence of general relativity. Aside from $e_{\rm max}$, none of the terms in these criteria depend on the parameters of the outer perturber inducing the EKL oscillation.} In other words, to evaluate whether GR is important when considering EKL oscillation from an outer companion, we only need to consider the eccentricity it induces on $m_2$. This is shown in Figure \ref{fig:grsuscept}, where we depict a population of 2-planet systems, with the diagonal lines indicating different choices of $e_{\rm max}$. Thus, potentially highlighting that for a {small but not insignificant} fraction of 2-planet population {(4-5\%)}, GR plays an important role in stabilizing the system.

Figure \ref{fig:grsuscept} depicts known two-planet systems, adopted from the {\tt exoplanet archive} \citep[][]{exoplanetarchive}, in the parameter space defined by the plane ${a_1^2}/{a_2}$-${M^2}/{m}$, motivated by Equation (\ref{eq:criterion}). We have removed systems orbiting binary stars{, systems without measurements of stellar mass }or where one or both of the two planets lack mass measurements. Overplotted are the boundary lines between GR dominating and Laplace-Lagrange dominating as a function of $e_{\rm max}$. To first-order, $e_{max}$ represents the maximum eccentricity that would be obtained within a few EKL or Laplace-Lagrange cycles in the presence of GR precession, which is straightforward to calculate \citep[e.g.,][]{Dan,Liu+15}. The value of $e_{max}$, can be extremely high, even in the presence of GR precession \citep[e.g.,][]{Naoz+12GR,Tey+13,Li+14coplaner}.  We note that at values of $e_{max}>0.8$ much of the parameter space lies below the transition curve, meaning that if $e_{\rm max}$ is high, GR is making a significant contribution to the stability of a non-negligible portion of the parameter space. It is worthwhile to point out that the Laplace-Lagrange formalism is an expansion in small eccentricities. 
However, even when higher orders in eccentricity are included, the combination of the GR and Laplace-Lagrange precessions can accurately predicate the stability criterion, as shown in \citet{wei+21}. {Note that other processes, especially when high mutual inclinations, and high eccentricities may exist, and higher orders of the approximation can affect the system \citep[e.g.,][]{Naoz16}.}

\section{An Inner Companion}\label{sec:Inner}

Most exoplanet detection methods are  sensitive to short orbital periods. For example, the Kepler telescope's data pipeline could detect planets on a $1.5$ year orbit at the longest (because of the pipeline's requirement of $3$ or more transits). Furthermore, Kepler's methods are insensitive to planets with radii smaller than about $0.5\, R_\oplus$. {For Earth-mass planets it is also insensitive to whose with orbital periods longer than about $150$ days} \citep[see Fig 1 of][]{Christiansen+17}.\footnote{Stellar and especially stellar activity may push this limit up and down. For example, the $0.354\, R_\oplus$ planet Kepler 37b was discovered in 2013 orbiting an especially quiet star \citep[e.g.,][]{Barclay+13}.} Thus, if two planets were detected in a system, an inner planet may still be hidden there with a radius smaller than $R_\oplus$. We now investigate the stability of such a planet under perturbations from its companions.

\subsection{The Role of GR - Stability Criterion for Inner Companion}
\label{sec:GRInner}

We now consider systems where a small (Super-Earth or lighter) planet ($m_{\rm in}$) orbits within one or more other 
planets with small mutual inclinations and arbitrary masses (see Figure \ref{fig:cartoon}). In this scenario, without an outer companion, the EKL mechanism is not pronounced, and the system is non-hierarchical, but significant eccentricities up to $\sim 0.4$ can still be driven on the super-earth by Laplace-Lagrange effects  (see also Section \ref{sec:LLGR} for example).

Depending on the orbital configuration, this could fatally disrupt this planet. However, if GR precession occurs on a shorter timescale than Laplace-Lagrange excitation, then the eccentricity that can be excited drops significantly. Because we are now concerned with the stability of the innermost planet, we must use the GR precession timescale of the inner planet, $\tau_{\rm GR,in}$ and require stability against Laplace Lagrange excitations from the further-orbiting $m_1$ on $m_{in}$. 
Note that in Section \ref{sec:OuterGR}, we consider the Laplace-Lagrange interactions between $m_1$ and $m_2$, where the latter is influenced by a distant companion (with an arbitrary geometry). Here we consider a different structure, where $m_1$ and $m_2$ are perturbing a hidden planet $m_{\rm in}$ (See Figure \ref{fig:cartoon} for an illustration of the system). We wish again to compare the Laplace-Lagrange and GR timescales as we did in Eq (\ref{eq:criterion}) to investigate stability. However, the new structure changes the semi-major axis dependence of the Laplace-Lagrange timescale (see Eq (\ref{eq:tau_LL_pm}) for details), so a new criterion must be derived. The Laplace-Lagrange timescale acting on $m_{in}$ from the perturbations of $m_1$ is given by Eq.~(\ref{eq:tau_LL_pm}): 
\begin{equation}\label{eq:tauLLin1max}
    \tau_{\rm LL,in1, max} \approx \left[\frac{7 n_{in} m_1}{4\pi M_\star} \left( \frac{a_{in}}{a_1} \right)^3 \right]^{-1} \ ,
\end{equation}
\begin{equation}\label{eq:tauLLin1min}
    \tau_{\rm LL,in1, min} = \left[\frac{n_{in} m_1}{4\pi M_\star} \left( \frac{a_{in}}{a_1} \right)^2 \left(f_{\psi}+{f_{2\psi}} \right) \right]^{-1} \ .
\end{equation}
Thus, the stability criterion of $\tau_{\rm GR,in}< \tau_{\rm LL,in1}$ yields {(using $\tau_{\rm LL,in1,max}$)}:
\begin{equation}\label{eq:InnerGRCrit}
    \frac{a_{\rm in}^4}{a_1^3} \leq \frac{6 G M_\star^2}{7 c^2 (1-e_{\rm max}^2) m_{\rm 1}} \ ,
\end{equation}
and using $\tau_{\rm LL,in1,min}$)
\begin{equation}\label{eq:InnerGRCritMin}
    \frac{a_{\rm in}^3}{a_1^2} \left(f_{\psi}+{f_{2\psi}} \right) \leq \frac{6 G M_\star^2}{c^2 (1-e_{\rm max}^2) m_{\rm 1}} \ ,
\end{equation}
where $e_{\rm max}$, is the maximum eccentricity achieved by $m_{\rm in}$ due to Laplace Lagrange resonance.

\subsection{Example of Resonances in Laplace-Lagrange Secular Evolution}\label{sec:LLGR}
The Laplace-Lagrange secular model has an analytical solution, where the eccentricity and inclination are, as functions of time, algebraic functions of sines/cosines of eigenfrequencies \citep[p.277-278]{Murray+00book}. {Therefore, there exist some configurations of planets where these eigenfrequencies approach one another in value, driving large eccentricities and inclinations in what we call a "Laplace-Lagrange Resonance"}. {Because the Laplace-Lagrange model assumes eccentricities remain low, and at these resonance points it can predict extreme eccentricity values, the assumptions implicit in Laplace-Lagrange formalism may be violated in these configurations. }Thus,  $n$-body simulations may result in higher but not necessarily unity eccentricities. 
Given two fixed planets, the semi-major axis values where a third planet would be in resonance are not affected by the masses of the two fixed planets. The relative masses of the planets do, however, affect the eccentricity the resonance predicts: if a planet is in resonance and is significantly less massive than its companions, its eccentricity will be excited to larger values.

\begin{figure}
 	\includegraphics[width=\columnwidth]{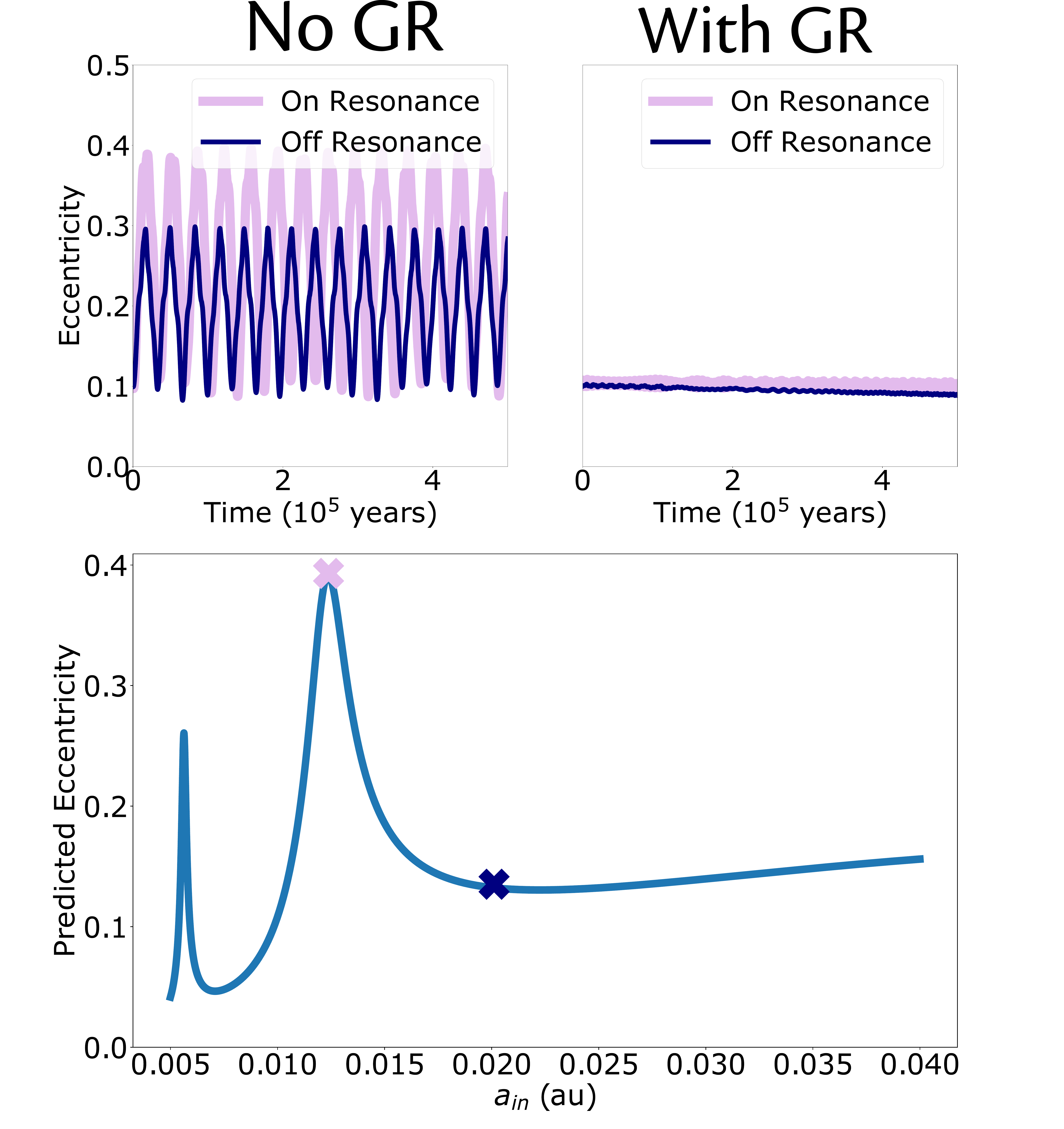}
     \caption{An Earth-mass planet was inserted within HD 15337's two orbits at a semi-major axis of peak resonance, $0.0125$ au (marked in light pink), and once off-resonance at $0.02$ au (marked in dark blue) both with an initial eccentricity of $e_{\rm in} = 0.01$ and integrated forward $50,000$~years using the $n$-body code {\tt MERCURY}. HD 15337 is a $0.9\,M_\odot$ star that hosts two known planets with $a_1=0.0522$ au, $m_1=7.51\,M_\oplus$, and $a_2=0.1268$ au, $m_2=8.11\,M_\oplus$ \citep[][]{Gandolfi+19}.  In the top panels the eccentricities both with (right) and without (left) including GR are plotted. In the bottom panel we show 
     the maximum eccentricity the Laplace Lagrange formalism predicts the planet should achieve over long timescales. The behavior of the eccentricity of the test planet is strongly effected by the predicted resonance, and in the presence of GR, the resonance is completely damped out.}
     \label{fig:resonancetest}
 \end{figure}

Angular momentum exchange between planetary orbits via the Laplace-Lagrange perturbations can result in eccentricity excitations even in the absence of an outer companion. In Figure \ref{fig:resonancetest}, bottom panel, we show the maximum eccentricity reached for an Earth-mass planet set at different separations as predicted from Laplace-Lagrange equations.
As a case study we consider the HD 15337 system, which is composed of  $M_\star=0.9\,M_\odot$ and has two planets with the following parameters:  $a_1=0.0522$~au, $m_1=7.51\,M_\oplus$, and $a_2=0.1268$~au, $m_2=8.11\,M_\oplus$ \citep[][]{Gandolfi+19}. In the bottom panel of Figure \ref{fig:resonancetest}, we show the maximum eccentricity predicted by Laplace-Lagrange's aforementioned analytical solution, as a function of a hypothetical inner planet's semi-major axis. This calculation is done in the $m_{\rm in}\rightarrow 0$ test-particle limit. Outside of this approximation, if $m_{\rm in}$ is set close to $m_1$ and $m_2$, the height of the resonance peaks decrease, but their locations as a function of $a_{\rm in}$ do not change.

We consider a test case of two specific points, one on-resonance, and one off-resonance. {Here "on-resonance" refers to a configuration where multiple frequencies in the $A$ matrix (\ref{eq:A_matrix}) are close or equal in value, resulting in large predicted eccentricities for one (or more) of the planets in the system. } For both of them we use the $n$-body code {\tt MERCURY}, integrating over $5\times 10^5$~yrs
\citep[][]{mercury6}. This version 
of {\tt MERCURY}, 
accounts for general relativity precession from the 1st post-Newtonian term, (M.~Payne private communication).  Here the Earth-mass planet eccentricity is initially set with $e=0.1$, and negligible mutual inclination with the known planets in the system. As depicted (top left panel in Figure \ref{fig:resonancetest}), over most of the parameter space, the eccentricity is excited up to about $0.2$--$0.3$, with fluctuation around that eccentricity possible in a configuration of Laplace-Lagrange resonance. On the other hand, (top right panel in Figure \ref{fig:resonancetest}), when the GR precession timescale is shorter than the typical Laplace-Lagrange precession, these eccentricity excitations is suppressed. In general, we find that regardless of whether this system is set up in a resonant configuration or not, when general relativity precession is included in the simulations, its timescale is short enough (by a factor of about $200$ in the resonant configuration) to dominate over Laplace-Lagrange and force the eccentricity to remain at its initial value. {We also note that general relativity precession is indeed damping out the Laplace-Lagrange resonance rather than simply shifting the eccentricity peak to another location.}

On the margins of the Laplace-Lagrange timescale/GR timescale equality, whether a resonance can destabilize a system where the GR timescale and the Laplace-Lagrange timescale differ by only a factor of unity is detail-dependent\footnote{ Similarly to the resonant-like behaviour that takes place when GR precession on a similar timescale to that of a quadrupole-level hierarchical three body system \citep[e.g.][]{Naoz+12GR}.}. However, we find that when the GR timescale is hundreds to thousands times shorter than the Laplace-Lagrange timescale, even resonances tend to be damped out.
 
\subsection{Numerical Comparison}\label{sec:num}
\begin{figure}
 	\includegraphics[width=\columnwidth]{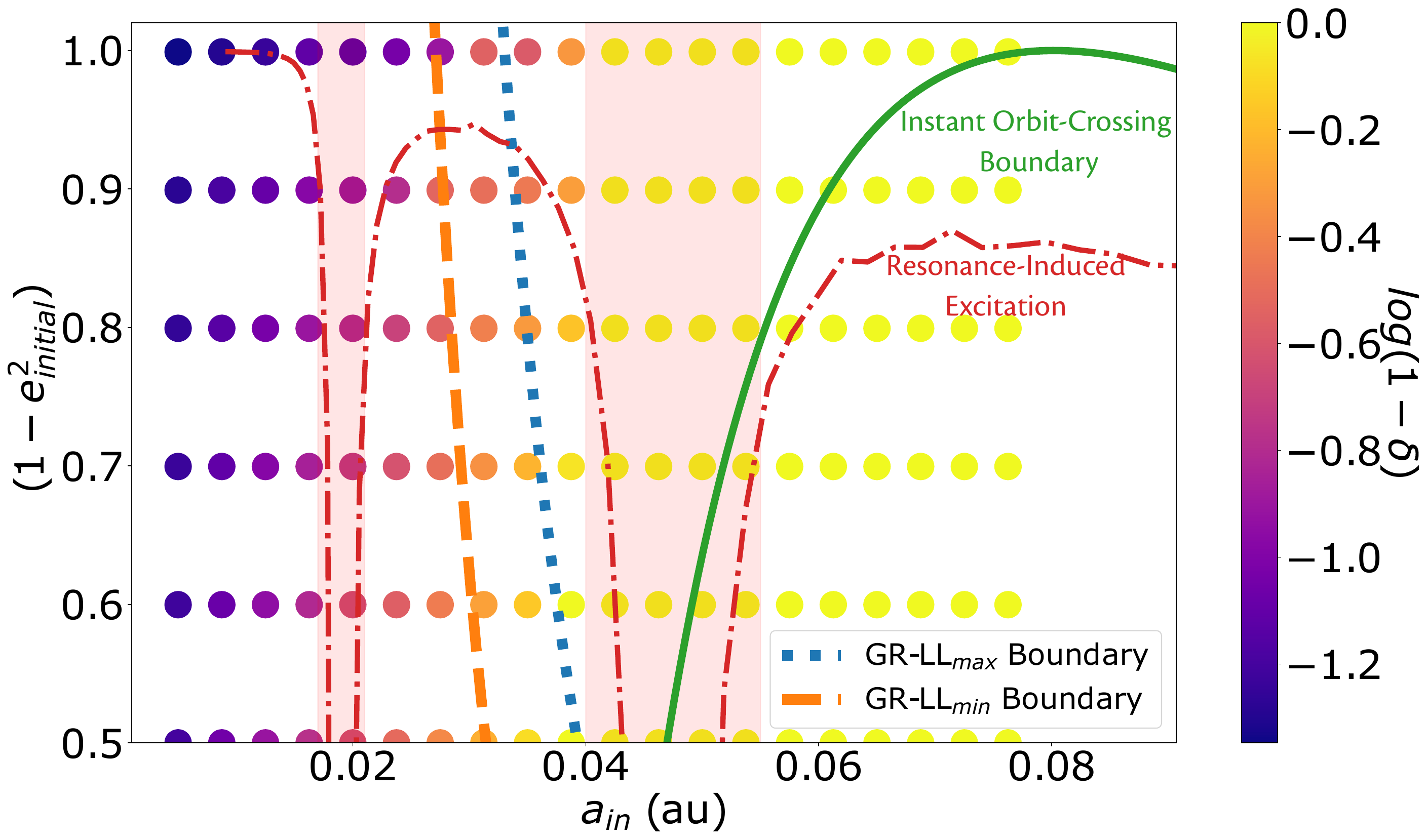}
     \caption{Stability map of a hypothetical-inner planet. In particular, we consider the HD 106315, composed out of two planets, HD 106315 b (c) with mass $12.6 \, M_\oplus$ ($15.2 \, M_\oplus$), semi-major axis $0.097$~au ($0.154 \,$au). We show the inner planet's initial specific angular momentum, as $1-e_{\rm initial}^2$ as a function of it's semi-major axis $a_{\rm in}$ (here, $e_{initial}$ spans the range from $0.01$ at the top, to $0.71$ at the bottom). The mass of the inner planet is $0.1\,M_\oplus$.  The points show the numerical integration of the system up to  $1$~Myr, where the color code represents how close the two orbits came to crossing over the integration according to Eq.~(\ref{eq:delta}). {The dot-dashed red line shows the Laplace-Lagrange prediction of the hidden planet's eccentricity, and regions where this eccentricity is especially high--the resonances--are shaded red.}   We over plot the GR-Laplace-Lagrange boundary according to Eqs.~(\ref{eq:InnerGRCrit}) and (\ref{eq:InnerGRCritMin}), thick, dashed orange, and dotted blue, respectively. We expect that to the left of these boundaries systems GR precession will suppress eccentricity excitations due to Laplace-Lagrange resonances, and to the right of these boundaries, resonances may lead to orbit crossing. Finally, the green line shows the orbit crossing limit.}
     \label{fig:paulplot}
 \end{figure}
 Next we test our stability criterion, Equation (\ref{eq:InnerGRCrit}) for the inclusion of a hidden inner planet against secular simulations. We utilize the Gauss averaging method to integrate the system, while including 1st Post Newtonian GR precession \citep[c.f.][]{wei+21}. In this approximation the line density of an orbit is inversely proportional to the velocity \citep[e.g.,][]{Touma+09}. This type of analysis was shown to be very efficient in calculating long-term evolution \citep[e.g.,][]{Nesvold+16,Denham+19,wei+21,Michtchenko+04}. The secular code allows us to study the system for long timescales as we note that the secular code agrees with the $n$-body code {\tt MERCURY},  \citep[][]{mercury6} with general relativity (M.~Payne private communication). 
 
 As a case study, we consider the HD 106315 system \citep[]{Zhou18,Livingston+18}, which consist of a 1.024~M$_\odot$ star and two known planets: HD 106315 b (c) with mass $12.6 \, M_\oplus$ ($15.2 \, M_\oplus$), semi-major axis $0.097$ au ($0.154$ au), and eccentricity $0.093$ ($0.22$) \citep[e.g,][]{Barros+17,Mayo+18,Crossfield+17,Rodriguez+17}. We consider a possible hypothetical inner planet with $m_{\rm in}=0.1 M_\oplus$  and explore the stable configuration of the system.   
 
 In Figure \ref{fig:paulplot}, we depict the hidden inner most planet's initial specific angular momentum $(1-e_{\rm initial}^2)$, as a function of its semi-major axis $a_{\rm in}$). Following \citet{Denham+19}, we quantify how close the orbits grow over the course of the simulation as:  
 \begin{equation}\label{eq:delta}
    \delta = \frac{a_1(1-e_1) - a_{\rm in}(1+e_{in, max})}{a_1-a_{\rm in}} \ ,
\end{equation}
shown through the coloring of the dots. {We highlight the locations of the Laplace-Lagrange resonances (the shaded regions), and the Laplace-Lagrange predicted eccentricity at all $a_{in}$ values (the dot-dashed red line)}.   We overplot our stability criterion, from Equation (\ref{eq:InnerGRCrit}) as the dotted thick line as well as a modified version of the criterion where we use the minimum Laplace-Lagrange timescale instead of the maximum (dashed line, see Eq.~(\ref{eq:InnerGRCritMin})). 
To the right of the this boundary where Laplace-Lagrange dominates, we see that the secular resonance on the inner planet is enough to drive orbit crossing ($\delta \rightarrow 0$), but on the left side of the boundary, where GR dominates, the secular resonance does not induce any additional eccentricity---even when initial eccentricities are set to be as high as $1-e_{init}^2 = 0.5$. We also see that the difference between using the minimum Laplace-Lagrange timescale instead of the maximum as we do in Equation (\ref{eq:InnerGRCrit}) does not make a significant difference in the location of the boundary between Laplace-Lagrange and GR dominance.

To the right of the large resonance we can see in the low initial eccentricity runs that although they are beyond the resonant region, enough eccentricity is excited to achieve orbit crossing. This arises from the structure of the resonances. In systems with exactly two planets, where there are two resonances within the orbit of the closest-in planet, typically the region between the second resonance and $a_1$, has a maximum eccentricity higher than the region inward to the first resonance (as also shown in Figure \ref{fig:paulplot}, dot-dashed line, note that this effect is less pronounced in Figure \ref{fig:resonancetest}, but still present). This behavior can drive orbital crossing even outside the resonant region.

\subsection{Observational Implications}
  \begin{figure}
 	\includegraphics[width=\columnwidth]{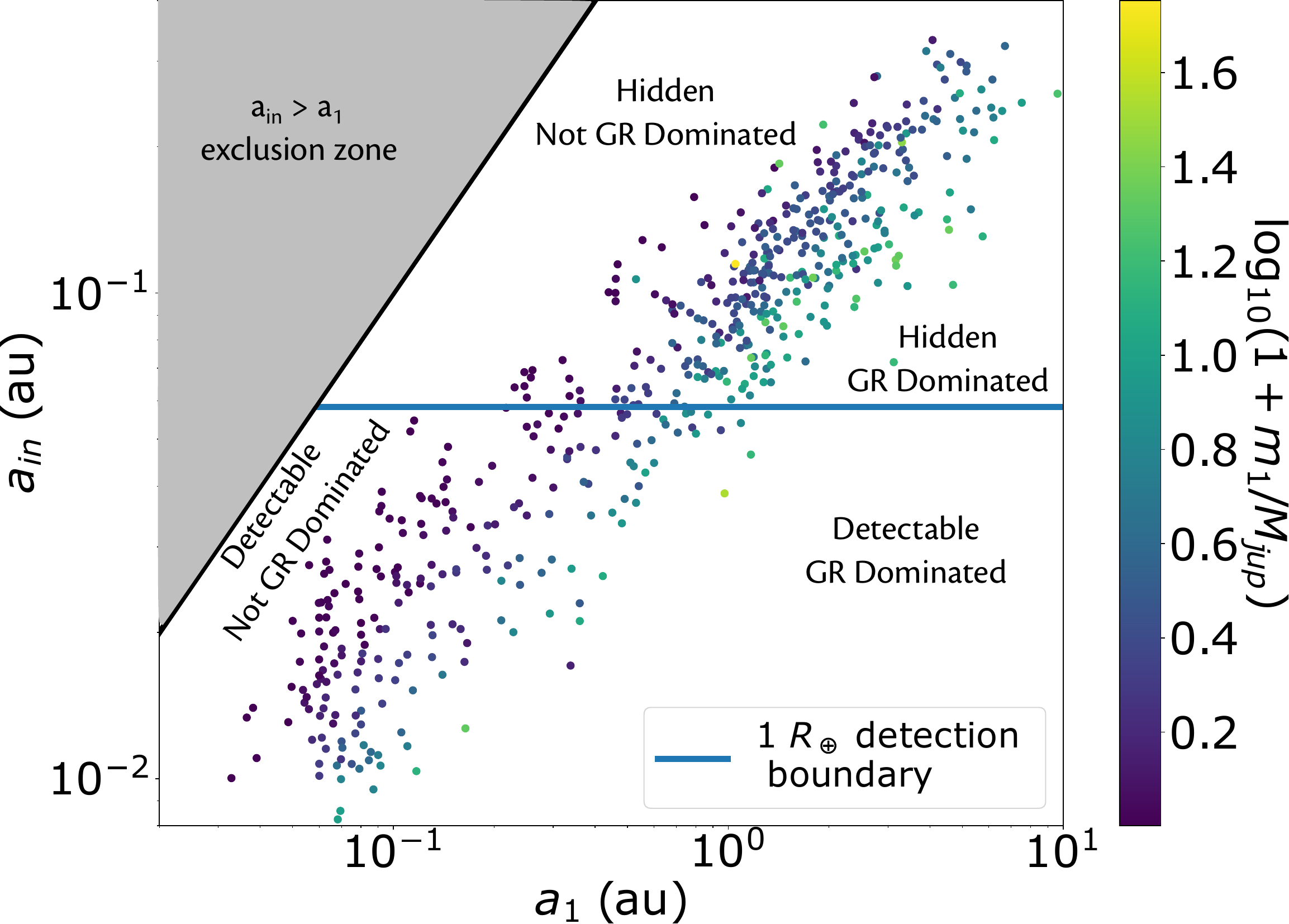}
     \caption{The relationship between the semi-major axis of the shortest observed orbit in a system ($a_1$) and the largest possible semi-major axis a hypothetical inner planet could have while still being GR dominated (see Eq.~(\ref{eq:InnerGRCrit})). Each dot represents a known main sequence star with either two or one known planets. The color represents the mass of the known planet $m_1$ in log. Overplotted is the boundary that separates the region where a $1\,R_\oplus$ inner planet would be detectable, via transit, according to Eq.~(\ref{eq:det}) if orbiting a sunlike star. Closer orbits (smaller $a_{\rm in}$) are more likely to be detected via transits and RVs.  The region of parameter space where $a_{in} > a_1$ has been grayed out as it is excluded by assumption. The inner planets were assumed to share Earth's mass and radius.}
     \label{fig:maxallowedsemi}
 \end{figure}

Figure \ref{fig:maxallowedsemi} depicts the largest possible semi-major axis ($a_{in}$) that a hypothetical, earth-analogue inner planet could have while still remaining GR-dominated for an observed system. We show this as a function of the semi-major axis of the closest-in observed planet in the system ($a_1$).  We limit the sample to planets orbiting main-sequence stars with two or fewer known planets. The points' colors represent the mass of the known planet, $m_1$. Below the points represents regions of parameter space where an inner planet would be GR dominated, above---Laplace-Lagrange dominated. 
For each star, a boundary exists where transiting planet can be detected (see Eq.~(\ref{eq:det})).

We use a simplified model of the Kepler mission pipeline's candidate confirmation requirement that the Multiple Event Statistic (MES) must be greater than $7.1$ \citep[]{2012PASP..124.1279C}. The MES is given approximately by
\begin{equation}\label{eq:det}
    \text{MES} \approx \left(\frac{R_{p}}{R_{\star}}\right)^2 \frac{1}{\textit{CDPP}_{\textit{eff}}} \sqrt{N_{t}} \ ,
\end{equation}
where $R_\star$ ($R_{p}$) is the stellar (planet) radius, $\textit{CDPP}_{\textit{eff}}$ is the effective stellar noise in parts per million, and $N_t$ is the number of transits the planet makes over the observing period. While this boundary depends on the stellar radius, as a proof-of-concept, here we over plot the  semi-major axis $a_{in}$ beyond which an earth-radius planet would be undetected, for a $1$~R$_\odot$\footnote{Note that the stellar radii of the observed systems in Figure \ref{fig:maxallowedsemi}, have a typical radius of $1$~R$_\odot$, up to about a factor of $2$. }.

The dispersion around the trend reveals that among the set of main sequence stars around which planets have been found, the widest orbit an inner planet can have and remain GR dominated is determined by $a_1$ to within about an order of magnitude.{ Importantly, \textit{in this case,} we need only information about one known planet in order to mark the locations in parameter space where a currently undetected planet may be hidden as is done in Figure \ref{fig:maxallowedsemi}.} Therefore, by needing very little information as depicted in Figure \ref{fig:maxallowedsemi}, the number of systems to which this criterion is applicable is maximized.

For a particular system, the area above the detection boundary and below that system's associated point, marks the region of parameter space that is both hidden from current observation and is GR dominated. {As we show in Figure \ref{fig:paulplot}, the GR dominated regime stabilizes such planets against Laplace-Lagrange eccentricity excitation. This Hidden and GR Dominated region of parameter space is more stable than non-GR dominated regions and more likely to contain planets than unstable regions}. {Notably, higher values of the known planets mass $m_1$ increases the strength of Laplace-Lagrange interactions between it and the inner planet. Thus, the largest allowable $a_{\rm in}$ where the inner planet is GR dominated must shrink as $m_1$ increases.}

Notably, higher values of the known planet's mas $m_1$ make the largest GR dominated $a_{in}$ in their systems smaller, thus shrinking the Hidden+GR Dominated portion of parameter space. Among systems where $m_1$ is less than a Jupiter mass, this region is quite a bit larger.

\section{Kepler-56: Proof of Concept (Outer and Inner Companions)}\label{sec:Kepler56}

Kepler-56 is a $1.32\,M_\odot$ red giant branch star, hosting three known planets \citep[e.g.][]{huber+13, otorB+16,Lissauer+11,hadden+14}. These planets feature a significant hierarchical structure---the semi-major axes of the three planets are $0.103$, $0.165$ and $2.16$ au, with masses of $0.07\,M_{J}$ and $0.57\,M_{J}$ for the inner planets, respectively. The inner two planets exhibit large obliquity with respect to the star's spin axis \citep{huber+13}. The outer planet is non-transiting, with an $M\sin i$ value of $5.61\,M_{J}$ \citep[e.g.][]{otorB+16}. The two inner planets have small eccentricities ($<0.05$), indicating that EKL excitations from Kepler-56d are not destabilizing this system, but can result in the large spin-orbit misalignment \citep{Li+14Kepler}. The outer planet has eccentricity of $\sim 0.2$ \citep[][]{otorB+16}. Indeed, applying the criterion defined in \citet{Denham+19}, we find that the critical eccentricity of Kepler-56 d above which we expect large eccentricity fluctuation of these planets is $\sim 0.83${, much larger than its reported eccentricity}.

Using Eq.~(\ref{eq:InnerGRCrit}), we can constrain a hypothetical Earth-radius planet in Kepler-56 to lie at less than $\lsim 0.07$~au, depending on this planet's initial eccentricity. However, we find that a Mercury-radius planet with sufficient eccentricity can remain GR dominated at orbits wider than $0.08$~au\footnote{We note that the stability limits uses Eq.~(\ref{eq:approx}), which is valid when  $a_1/a_2<0.9$, which {somewhat} limits the extent of the use of the equation.} but probably cannot be wider than $0.088$~au, which represents about $5$  mutual Hill radii from Kepler-56 b \citep[the spacing that can yield stability against planet-planet scattering][]{Chatterjee+08}.
From Eq.~(\ref{eq:det}) we see that, such a planet could transit and remain undetected at orbits comparable to its Roche limit with the host star. 

\begin{figure*}

    \centering
    \includegraphics[width=\linewidth]{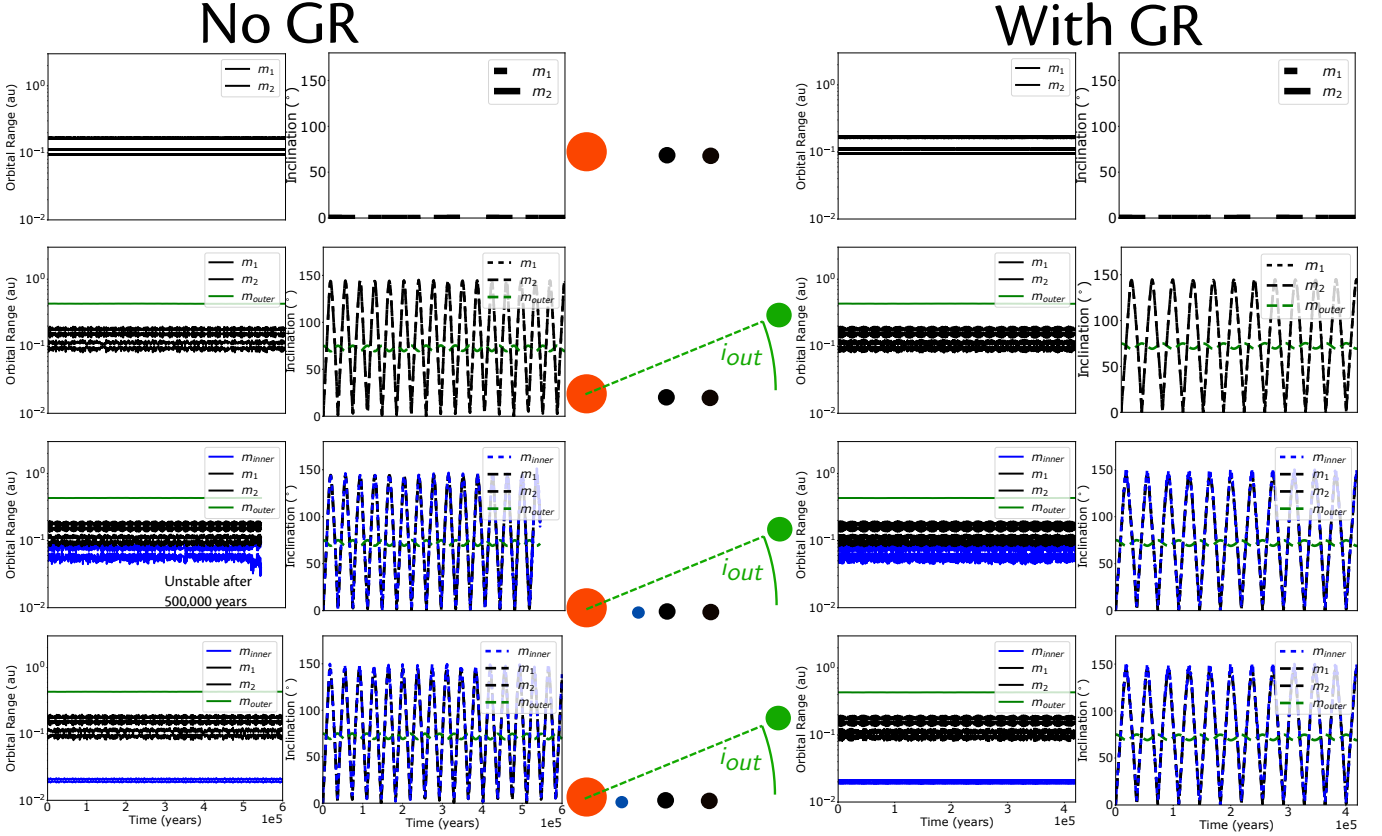}
    \caption{Orbital outcomes of our {Kepler 56-like} system in 4 configurations, with all 4 configurations run once without GR and once with GR. Top to bottom (as indicated by the cartoons in the middle) are the eccentricities and inclinations of the constituent planets when (i) The outermost planet is removed (ii) The outermost planet is included (iii) An Earth-mass planet is inserted into the system close to the innermost planet (iv) An Earth-mass planet is inserted into the system close to the star. Kepler 56 is a $1.32\,M_\odot$ star that hosts three known planets with $a_1=0.1028\,$au, $m_1=0.58\,M_{J}$, $\,e_1=0.04$, $a_2=0.165\,$au, $m_2=0.57\,M_{J}$, $\,e_2=0.01$, and $a_3=2.16\,$au, $m_3=5.61\,M_{J}$ \citep[][]{huber+13,otorB+16}. We set $e_3=0.8$ rather than its measured value of about $0.2$ for the purposes of proof-of-concept. The outer planet's inclination was set to $i_{\rm out}=75^\circ$. The inner planet is one Earth-mass and starts the integrations with negligible eccentricity}
    \label{fig:panelplots}
\end{figure*}

Because Kepler-56 contains a known outer planet, we observe the behavior of the system after the insertion of a hypothetical inner planet in Figure \ref{fig:panelplots}. We simulate the system both with and without GR in four configurations depicted in the middle column of Figure \ref{fig:panelplots}. {The first configuration has no inner planet, and the known outer planet has been removed. The second brings back the outer planet, and its eccentricity has been increased to $e_{\rm out} = 0.8$ to exaggerate its effect for proof of concept purposes.}  The third and fourth configurations introduce a hypothetical inner companion first at an orbit close to its neighboring planets where the GR and Laplace-Lagrange timescales are roughly equal, and second at an orbit close to the host star where the GR timescale dominates. For each configuration we plot the inclination of all the planets and their "orbital range" or the location of each planet's periapsis and apoapsis as functions of time. If one planets periapsis overlaps with another's apoapsis, we conclude the system will undergo an eventual orbit crossing and is likely unstable. The EKL timescale is smaller than both in all configurations, but because of Laplace-Lagrange and GR having weak influence over inclinations, the outer planet still causes large, synchronized inclination shifts. Broadly, we find that this system is quite stable even with the outer planet's eccentricity artificially increased. However, we find that in the third configuration, the hypothetical planet does undergo orbit crossing after 500,000 years, but only if GR is neglected. Because the outer known planet's eccentricity has been measured to be about $e_{\rm out} = 0.2$ \citep[][]{otorB+16}, the system is even more stable than these plots show. {This modified system is a valuable test case to show how a multiplanet system can be stable in the presence of an extreme companion (where we have adopted a higher eccentricity for the outer planet to more clearly demonstrate its effect) and a possible inner planet. We further demonstrate in Figure \ref{fig:panelplots} third row that GR precession extends the stability regime.}

\section{Discussion}\label{sec:diss}
Our current exoplanet detection methods are less effective at detecting smaller planets and far-orbiting planets than larger or closer-in planets. Already, for example, it is estimated that about $50\%$ of stars may host Jupiter-like planets at wide separations \citep[$5-20$~au]{Bryan+16}. In general it was suggested that a population of giant planets often host one or more inner planets \citep[e.g.,][]{Knutson+14K,Konopacky+2016,Zhu+18,Bryan+16,Bryan+19}.  Further, some  close-in small planets have already been observed \citep[e.g.,][]{2014ApJ...787...47S,2020ChA&A..44..283X,2017AJ....154..226D}. 
Here we study the stability of known, at most two-planet systems, hosting either a far away companion or a inward small planet.

We have quantified a condition under which the addition of a new planet into a known system does not destabilize the system, considering both inner and outer hypothetical planets.  A far away companion, may excite the inner system's eccentricities, via the EKL mechanism driving them to instability. {However, Laplace-Lagrange interactions between neighboring planets can tend to stabilize the system} \citep{Pu+18,Denham+19}. Recently \citet{wei+21} showed that GR precession expands the stability regime. {Here we consider all systems with more than one observed planet, and constrain the regimes where GR precession expands the stability of the system against EKL excitation more than Laplace-Lagrange on its own (depicted in Figure \ref{fig:grsuscept})}. This regime depends on the maximum eccentricity excited via EKL form the far-away companion which often can be extremely high \citep[e.g.,][]{Li+14coplaner}. 

Further, we developed analytic criteria for the stability of an inner planet (see Equations (\ref{eq:InnerGRCrit}) and (\ref{eq:InnerGRCritMin})). In this case, motivated by {\tt Kepler} sample of ``peas in the pod'' \citep{Weiss+18Peas}, we assume a low-inclination planet inwards of a known planetary system. {Laplace-Lagrange resonances can drive this inner planet to high eccentricity values.} General relativity precession can suppress these eccentricity excitations, as highlighted in Figure \ref{fig:resonancetest}. In some cases, Laplace-Lagrange resonances may drive the system to instability, i.e. orbital crossing. We show that our analytical criteria (Equations (\ref{eq:InnerGRCrit}) and (\ref{eq:InnerGRCritMin})) are in good agreement with numerical calculation, as depicted in Figure \ref{fig:paulplot}.     

Using our analytical criterion we examined all systems with either one or two planets and estimated the largest possible, stable semi-major axis of a hypothetical planet (see Figure \ref{fig:maxallowedsemi}). A planet could be hidden in a system that is only Laplace-Lagrange dominated, but if a resonance exists, some locations my be unstable. On the other hand, the GR precession dominated regime will suppress Laplace-Lagrange resonances, making it a {stable} location for short period planets to exist (right side in Figure \ref{fig:maxallowedsemi}). 

We note that stellar oblateness (also known as J2) can induce nodes precession of an ultra short planet. It was suggested that this precession  may even result in inducing small mutual inclination between multi-planet systems \citep[$\lsim 40^\circ$~e.g.,][]{Becker+20,Li+20,Schultz+21}. As highlighted in \citet{wei+21}, stability analysis can be done for wide variety of short-range forces. Thus, it is straightforward to estimate the induced precession due to the stellar rotation. We find that for a Sun-like star system, assuming a Sun-like rotation rate ($\sim 25$~d period), the oblateness-induced precession timescale is much longer over most part of the parameter space (e.g., in Figures \ref{fig:paulplot} and \ref{fig:maxallowedsemi}). However, we note that G-type stars probably slow their spin during their evolution due to magnetic braking \citep[e.g.,][]{Dobbs+04}. Thus, it may be that the oblateness-induced precession was more significant compared to GR precession, for young, fast spinning, stars.

Finally, we highlight the relevance of GR precession by considering the system, Kepler-56 \citep{huber+13}, and starting with two-planet systems, we systematically add planets (see Figure \ref{fig:panelplots} while comparing to evolution with and without GR. We first (second panel) add Kepler-56d, although for a proof-of-concept we adopt a higher eccentricity for it, $0.8$. The system indeed is stable to even such high eccentric companion \citep[e.g.,][]{Denham+19}. We then add a smaller planet inward to Kepler-56b\footnote{Note that such a planet will not survive Kepler-56 radial expansion as it continues to evolve \citep{Li+14Kepler}, but it may result in some ejections from the star \citep[e.g.,][]{Stephan+20}. }, and show two examples. One of which GR precession suppress eccentricity excitations, due to a resonant location, and another one, off-resonance. {As shown in Figure \ref{fig:panelplots}, a combination of the Laplace-Lagrange and GR precessions stabilize an extreme Kepler-56 like system.} In particular, an even  more eccentric outer companion than the one observed can still keep  the system stable \citep[consistent with][]{Denham+19}. Moreover, the system can hide a hypothetical small companion, inward to Kepler 56b, and remain stable.   

The aforementioned examples highlight the application of our analytical stability criteria in constraining the possible configurations of hidden planets. {In particular, GR precession induced by the star can increase the stabilization of a multi-planet system against perturbation from a far away companion.} Further, we showed that GR precession could also stabilize an ultra-short period hidden planet, even in the presence of resonances induced by a multi-planet system.

\section*{Acknowledgements}
\begin{acknowledgments}
We thank Matt Payne for his {\tt Mercury} version with GR precession.We thank Gongjie Li for useful discussion and comments on an earlier version of the paper. Likewise, we would also like to thank Juliette Becker and Peter Plavchan, and Jon Zink.
This research has made use of the NASA Exoplanet Archive, which is operated by the California Institute of Technology, under contract with the National Aeronautics and Space Administration under the Exoplanet Exploration Program.
T.F and S.N. acknowledge partial support from the NSF through grant No. AST-1739160. S.N. thanks Howard and Astrid Preston for their generous support.  L.W and S.N acknowledge the support from the Cross-Disciplinary Scholars in Science and Technology (CSST) program of University of California, Los Angeles.
\end{acknowledgments}

\bibliographystyle{mnras}
\bibliography{paper} 

\end{CJK*}
\end{document}